# Deep learning enabled superfast and accurate M² evaluation for fiber beams

YI AN,[1] JUN LI,[1] LIANGJIN HUANG,[1,2] JINYONG LENG,[1] LIJIA YANG,[1] AND PU ZHOU[1,3]

[1]*College of Advanced Interdisciplinary Studies, National University of Defense Technology, Changsha 410073, China*
[2]*hlj203@nudt.edu.cn*
[3]*zhoupu203@163.com*

**Abstract:** We introduce deep learning technique to predict the beam propagation factor M² of the laser beams emitting from few-mode fiber for the first time, to the best of our knowledge. The deep convolutional neural network (CNN) is trained with paired data of simulated near-field beam patterns and their calculated M² value, aiming at learning a fast and accurate mapping from the former to the latter. The trained deep CNN can then be utilized to evaluate M² of the fiber beams from single beam patterns. The results of simulated testing samples have shown that our scheme can achieve an averaged prediction error smaller than 2% even when up to 10 eigenmodes are involved in the fiber. The error becomes slightly larger when heavy noises are added into the input beam patterns but still smaller than 2.5%, which further proves the accuracy and robustness of our method. Furthermore, the M² estimation takes only about 5 ms for a prepared beam pattern with one forward pass, which can be adopted for real-time M² determination with only one supporting Charge-Coupled Device (CCD). The experimental results further prove the feasibility of our scheme. Moreover, the method we proposed can be confidently extended to other kinds of beams provided that adequate training samples are accessible. Deep learning paves the way to superfast and accurate M² evaluation with very low experimental efforts.

## 1. Introduction

The characterization of laser beam quality is a concerning topic in laser fundamental and applied research. Over the past few decades, various parameters such as the beam propagation factor M², Strehl ratio, and power-in-the-bucket etc., have been proposed for assessing the beam quality [1–8]. Among them, M² parameter has received much attention and become the most popular metric, which compares the product of waist radius and divergence half angle of the beam under test to that of a fundamental Gaussian beam [4]. The International Organization for Standardization (ISO) has provided a standard M² caustic measurement procedure [9], which calculates the beam sizes at a range of positions near its waist so that the second-order moments of the beam and hence the M² value can be determined. This procedure is experimentally complex and relatively time-consuming (usually requires at least several minutes), restricting its application on the time-varying laser beam measurement. Thus, in terms of faster M² measurement, various methods have been proposed and demonstrated. For example, the motion-free variable-focus techniques utilize spatial light modulators [10] or liquid lenses [11], leading to a measuring time below 1 s. In addition, some single-shot schemes have also been reported, with which the beam cross sections can be imaged simultaneously on detectors using a distorted diffraction grating [12] or angled Fabry-Perot filter [13] and these schemes are tested by high power Nd:YAG lasers [12] or high power fiber lasers [13] respectively. Further, complex amplitude reconstruction methods using different kinds of interferometers [14,15] or just Charge-Coupled Device (CCD) [16] are proposed. The latter scheme [16] utilizes two identical CCDs to obtain beam intensity images at different defocused positions for wavefront reconstruction, making the time taken from the image acquisition of He-Ne laser beam to the $M^2$ value determination only about 0.5 s.

Actually, there is another class of $M^2$ determination approaches based on mode decomposition (MD). O. Schmidt et al. have utilized correlation filter technique to perform MD for the beams emitting from Nd:YAG laser and then $M^2$ is directly calculated according to the measured modal weights [17]. They also investigated the beam quality of fiber beams based on the same MD technique [18]. However, a direct calculation is replaced by a virtual caustic measurement (VCM) [18], which means that the free-space propagation of the fiber beam is simulated based on the MD results and then $M^2$ can be obtained. Besides the VCM method for fiber beams, direct calculation approach based on the electric field is proposed by H. Yoda et al. [19] without free-space propagating simulation, which indicates much less calculation cost. With Yoda's theory, the $M^2$ of the fiber beam is estimated based on the numerical MD results of the near-field beam intensity recorded by CCD, and the estimated $M^2$ agrees well with standard caustic measured value [20], which proves the accuracy of Yoda's theory [19]. Moreover, with the combination of the numerical MD and direct $M^2$ calculation theory, real-time $M^2$ estimation is achieved [21] and the processing rate reaches 9 Hz.

As a hot technique in recent years, the convolutional neural network (CNN) connects multiple units from layer to layer by linear or nonlinear operations [22–26], learning the complex mapping between different domains. CNN has extraordinary advantages on image processing. It can be used for not only classification but also regression problems with the help of different layers, which include convolution layers to extract proper features of the input image, pooling layers to reduce the size of feature maps, and fully-connected layers to have a deep understanding of the whole image. These layers allow a precise learning of the complex relationship between the input image and the output target. Therefore, CNN has been applied successfully in many fields of optics and photonics [27–30]. These applications exhibit the superb advantages of CNN than the conventional methods. In our previous work, we have utilized CNN to perform numerical MD for near-field fiber beam images and achieved accurate and fast MD results [30]. This inspired us that the $M^2$ estimation can also be successfully performed from a single-shot collected beam pattern by using the CNN.

In this paper, we have developed deep CNN for superfast and accurate $M^2$ prediction. Compared with traditional schemes [9–21], our approach is very economic, which only needs a CCD camera to obtain near-field beam patterns. The extraordinary time efficiency is another advantage of our scheme. Utilizing a trained CNN, the $M^2$ of the fiber beams can be determined in about 5 ms for one prepared beam pattern, which is potential to improve the processing rate from 9 Hz to 200 Hz for real-time $M^2$ evaluation on time-varying beams. Besides, different from traditional approaches, our scheme is of great robustness for imperfect beam patterns, e.g., noisy patterns, which greatly reduces the efforts to capture ideal patterns.

## 2. Methodology

### 2.1. Principle of the scheme

The principle of our scheme is illustrated in Fig. 1. The near-field beam intensity of the fiber beam is acquired by the CCD and then the beam pattern can be processed by the trained CNN to give the estimation of $M^2$ parameter along two orthogonal directions. To train the CNN, large amounts of near-field beam patterns and corresponding $M^2$-parameter need to be pair-collected either in experiments or simulations. A 4-*f* system consisting of two lenses is usually needed to collect the near-field beam pattern in the experiment, which will be introduced in Section 3.2. In our previous work [30], we have demonstrated that the simulated samples considering the real experiment condition are effective for the training of the prediction network, which is able to accurately analyze the experimental data. Here, the CNN training is also based on the simulated samples, which will highly reduce the complexity and improve the efficiency. The CNN training procedure is also illustrated in Fig. 1 for a better understanding of our scheme, the details of which will be introduced in the next section.

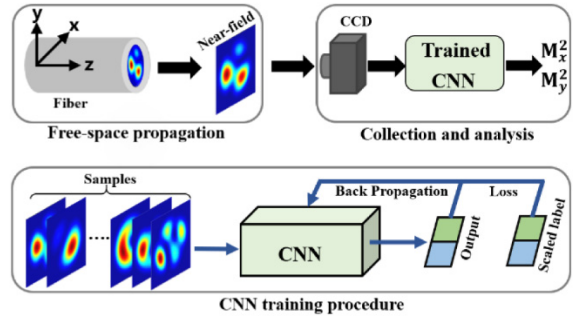

Fig. 1. An illustration of the scheme of deep-learning-based M² evaluation. The near-field pattern of the fiber can be easily recorded by a CCD and analyzed by a trained CNN.

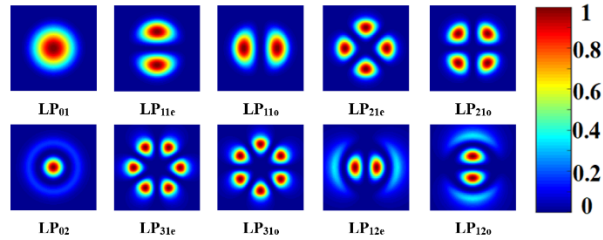

Fig. 2. The intensity profiles of the eigenmodes involved. The near-field patterns are generated through the superposition of eigenmodes.

To demonstrate the concept of our deep-learning-based $M^2$ prediction, a step-index fiber is taken as an example. Here, the fiber is assumed to work at 1064 nm with a core diameter of 25 μm and the NA of 0.08, a typical large-mode-area fiber for fiber lasers. The near-field electric field can be mathematically expressed as [31]

$$E(x,y,z_0) = \sum_{n=1}^{N} \rho_n e^{i\theta_n} \psi_n(x,y,z_0) \tag{1}$$

$$\sum_{n=1}^{N} \rho_n^2 = 1 \quad \theta_n \in [-\pi, \pi] \tag{2}$$

where $\psi_n(x,y,z_0)$ is the electric field of the $n^{th}$ eigenmode in the fiber with modal amplitude $\rho_n$ and phase $\theta_n$. The eigenmodes could be described by linearly polarized (LP) modes based on weak-guidance approximation [31]. This fiber can support as many as 10 eigenmodes, which can be arranged in the order of $LP_{01}$, $LP_{11e}$, $LP_{11o}$, $LP_{21e}$, $LP_{21o}$, $LP_{02}$, $LP_{31e}$, $LP_{31o}$, $LP_{12e}$, and $LP_{12o}$ modes. The intensity profiles of these eigenmodes are displayed in Fig. 2. Due to the degeneracy of the modes [31], 5 possible cases that the former 3, 5, 6, 8 or 10 modes propagating in the fiber are discussed respectively in our work.

Different sample images can be easily acquired by varying $\rho_n$ and $\theta_n$ in (1) with

$$I = |E(x,y,z_0)|^2 \tag{3}$$

As for the label, the beam quality factor along two orthogonal directions $M_x^2$ and $M_y^2$ can be calculated from Yoda's theory [19], which can be expressed as

$$M_k^2 = \sqrt{4B\sigma_k^2(z_0) + A^2}, k = x, y \tag{4}$$

The parameters in (4) are defined as

$$\sigma_k^2(z_0) = \iint (k - \langle k \rangle(z_0))^2 \, |E(x,y,z_0)|^2 \, dxdy \tag{5}$$

$$A_k = \iint (k - \langle k \rangle(z_0))(E(x,y,z_0) \times \frac{\partial E^*}{\partial k}(x,y,z_0) - c.c.) dxdy \tag{6}$$

$$B_k = \iint \left|\frac{\partial E}{\partial k}(x,y,z_0)\right|^2 dxdy + \frac{1}{4}\left(\iint \left[E(x,y,z_0) \times \frac{\partial E^*}{\partial x}(x,y,z_0) - c.c.\right] dxdy\right)^2 \tag{7}$$

where $\langle k \rangle$ is the center coordinate in $k$ axis, $E^*$ is the conjugate field of $E(x,y,z_0)$ and *c.c.* denotes the conjugate polynomial of the former one. Based on (4)-(7), theoretical values of $M_x^2$ and $M_y^2$ can be derived and utilized to train CNN.

The resolution of the generated pattern sample is set to 128 × 128 and the label of patterns is linearly scaled to [0, 1] by dividing a constant value for better training accuracy. This value is 3 for 3-mode and 5-mode cases or 4.5 for 6-mode, 8-mode and 10-mode cases separately, which is determined by the maximum of theoretical $M^2$ value [32].

## 2.2. CNN model

Our work is performed by a modified CNN from the VGG-16 model [22], which is a mature and accurate archtecture for image processing. The model is modified according to the input and output of the network. Concretely, the filter size of the first convolutional layer of VGG model is changed from 3 × 3 × 3 to 3 × 3 × 1, as our input is a single gray beam image. The dimension of the output vector of the last fully-connected layer is modified to two to ensure that the output vector size is equal to the label size. The Softmax layer of the origin VGG model is also replaced with a Sigmoid layer for our regression problem. Our CNN model can be divided into 7 blocks, as shown in Fig. 3. For the first five blocks, every block includes two or three convolutional layers and a max pooling layer. The ReLU activation layer after each convolutional layer is hidden in Fig. 3 for better illustration. The last two blocks are two fully-connected layers and the channel of them is set to 1024 and 2 respectively.

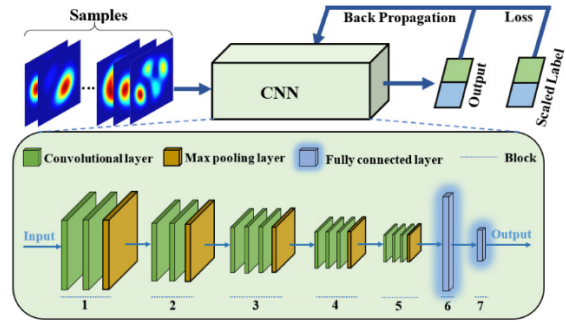

Fig. 3. The architecture of our CNN model, which includes 7 blocks.

The CNN learns to estimate $M^2$ parameter from the near-field beam intensity. During the training process, the input images are passed through the layers of the CNN and regressed into a 2-D vector in the forward propagations. We define the loss of our network as mean-square error (MSE) between the output and the scaled label vector. In the back-propagation stage, the parameters of the CNN are updated iteratively through stochastic gradient descent (SGD) based on the MSE loss.

When the network reaches convergence, it can be utilized for M² prediction. Taking a gray image as input, the CNN output a 2-D vector, from which the predicted M² value can be obtained by multiplying defined constant value mentioned in the label-scaling operation.

## 3. Results and discussion

All results reported in this paper are achieved on a desktop computer with an Intel Core i7-8700 CPU and GTX 1080 GPU. CNN is trained for the five cases (3, 5, 6, 8 or 10 eigenmodes involved) separately. The GPU is utilized to accelerate computing efficiency. We randomly generate 10000 images online in every training epoch, which means a time period. The learning rate is set to 0.01 in the first 20 epochs and 0.001 in the following epochs. The network gets convergence after 50 epochs and the total training time is about 2 hours for each case.

### 3.1. Analysis based on simulated beam samples

We use 1000 simulated beam profiles that are not contained in the training samples to evaluate the performance of the CNN for 5 cases involved respectively. The codes and data set of our scheme can be found on Github at https://github.com/anyi0924/M-2-estimation-through-deep-learning.git..

The effective beam propagation factor $M_{eff}^2$ can be defined as $M_{eff}^2 = \sqrt{M_x^2 \times M_y^2}$ [9], and this general parameter offers a simple and precise way to evaluate the accuracy of our scheme. For the $i^{th}$ testing sample, the prediction error (PE) can be shown as

$$PE(i) = \left| M_{eff,p}^2(i) - M_{eff,l}^2(i) \right| / M_{eff,l}^2(i) \tag{8}$$

where '$p$' denotes the predicted value and '$l$' denotes the label value.

We calculate the averaged PE of testing samples after every training epoch and the results are reported in Fig. 4. It can be found that the averaged PE decreases in the training process and finally becomes steady, indicating the CNN gets convergence. Then this trained CNN can be utilized for M² estimation, which only takes about an averaged time of 5 ms to give a prediction of $M_x^2$ and $M_y^2$ for a single prepared beam pattern of these five involved cases. The averaged PE with the pre-trained CNN can reach 0.4%, 1.3%, 1.6%, 1.8% and 2.0% for five cases respectively. It should be noted that these results are acquired based on the testing of near-field pattern samples. When the trained CNN is used to process the patterns at other imaging planes, the results will be not reliable since all of our training images are the near-field beam patterns. To obtain an accurate M² estimation value for the patterns at other imaging planes, the training has to be repeated with the corresponding samples.

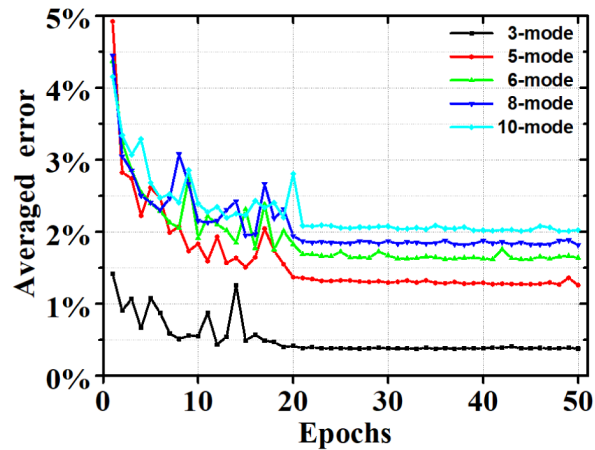

Fig. 4. Averaged prediction error as a function of epochs for 5 cases.

Besides the averaged error, the error distribution is another concerning issue of our scheme. To further observe the prediction performance of our scheme, we illustrate the PE distribution of testing samples of five cases respectively, as shown in Fig. 5. We set 10 fixed PE values, which are 1% to 10%, at the horizontal axis. Then the testing samples in every case with smaller PE than the fixed value are counted and the corresponding percentage is calculated as the ordinate value to match the 10 fixed abscissa values. To see the data points clearly, lines are added to connect them. From which we can find that for the 3-mode case the network can achieve a PE smaller than 2% for almost all samples while for the 10-mode case pre-trained CNN can achieve a PE smaller than 5% for about 95% samples.

Some typical results of our predicted $M_{eff}^2$ value and their corresponding labels are compared in Fig. 6 with the input pattern samples shown in the insets. The patterns A-J are selected from five cases respectively. The predictions and labels of $M_{eff}^2$ for these typical examples are very close, even for some very complex patterns, exhibiting the high accuracy of the trained CNN. The video of more input patterns as well as their corresponding labels and predicted values for $M^2$ in 5 cases is illustrated in Visualization 1, Visualization 2, Visualization 3, Visualization 4, and Visualization 5, respectively.

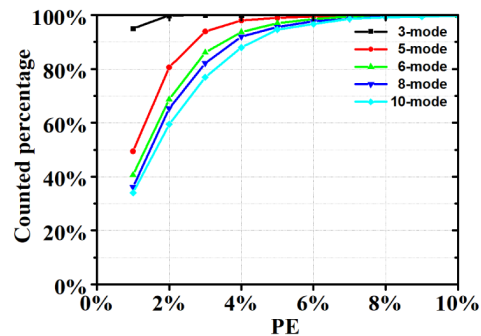

Fig. 5. The distribution of PE for testing samples in five involved cases. The samples whose PE smaller than fixed values are counted and the corresponding percentage are calculated.

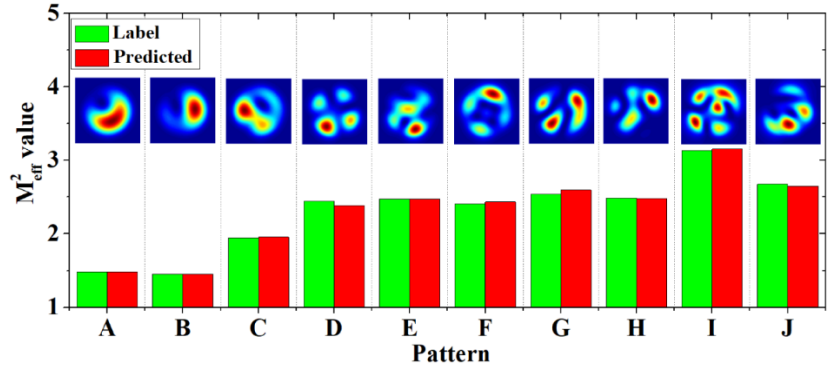

Fig. 6. The comparison of label and predicted $M_{eff}^2$ value for ten typical pattern samples selected from five cases respectively. A and B: 3-mode case; C and D: 5-mode case; E and F: 6-mode case; G and H: 8-mode case; I and J: 10-mode case.

Here we give more results about the predicted and ground truth $M^2$ value for a detailed view of our results. We arrange the ground truth of $M^2$ value of 100 testing samples in ascending order for five involved cases respectively, as shown in Fig. 7 with black spots. The corresponding predicted values represented by red spots are also plotted. Corresponding lines connect the black or red spots for better observation. From the illustration, we can find that the prediction fits the label very well for 3-mode and 5-mode cases while for the other cases these prediction values fluctuate in a very small range around the ground truth, showing the robust and high accuracy of our approach. Noted that when the mode number increases, the deviation between the predicted value and the labeled one becomes larger. To handle the more complex patterns, *e.g.*, patterns contain a multiple of 10 modes, larger number of layers is needed to learn the relationship between the patterns and $M^2$ parameter. Moreover, increasing training samples can also help enhance accuracy.

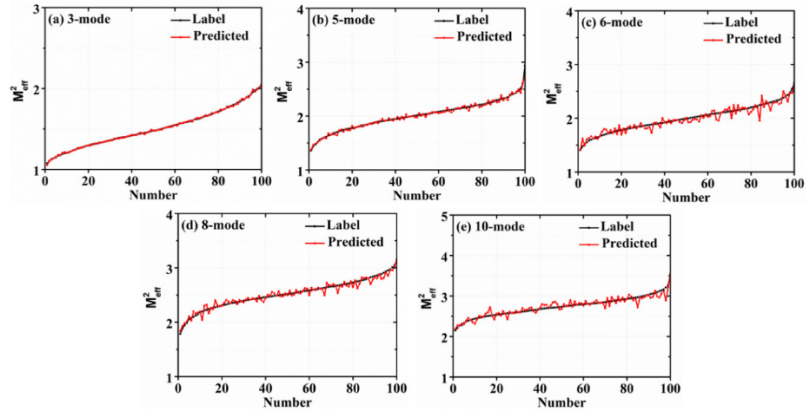

Fig. 7. Comparison of the label and predicted $M^2$ for five involved cases. (a) 3-mode case; (b) 5-mode case; (c) 6-mode case; (d) 8-mode case; (e) 10-mode case.

We also investigate the robustness of our CNN by adding noise to the input beam patterns. 1000 noisy testing samples under different noise intensity levels are prepared and fed into the trained CNN for each case. For the generation of noisy patterns, every pixel of clean testing pattern is multiplied by a factor, which equals to $1+N(0,1)\cdot\sigma$. Here $\sigma$ is defined as noise intensity and $N(0,1)$ is the standard normal distribution. The averaged PEs is calculated under different noise intensity levels, which are shown in Fig. 8. Some typical patterns with different $\sigma$ value (between 0 to 0.24) is also plotted on the top of Fig. 8 and it is reported that $\sigma$ value can hardly reach over 0.08 in most practical scenarios [29]. The PE is close to clean pattern inputs when $\sigma$ value is smaller than 0.08. When $\sigma$ increases, the averaged PE of 5 cases becomes lager, especially for 3-mode and 5-mode case, the reason of which might be that diversity of patterns offers a higher anti-noise ability when more eigenmodes are involved. Noticed that even when the noise intensity increases to 0.24, the averaged PE of 5 cases are still lower than 2.5%, which definitely proves the extraordinary robustness of our trained CNN.

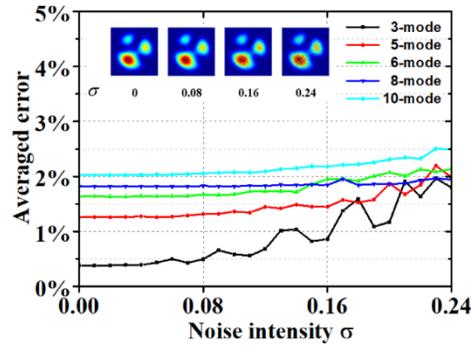

Fig. 8. The errors of $M^2$ prediction for noisy beam patterns. The typical noisy patterns with different noise intensity parameter $\sigma$ are shown on the top.

In the practical case, the resolution of CCDs always has several choices such as 1280 × 1024 pixels or 1024 × 768 pixels. The near-field pattern may appear on a small area of the CCD detectors so the valid resolution is much smaller. Noted that the resolution of image samples is set to 128 × 128 in our work, which simulates the valid area of the patterns recorded by a CCD camera. Different resolutions might influence the accuracy and time efficiency of our scheme. Here we also investigate the performance of our method with 96 × 96 and 160 × 160 pattern inputs for the 6-mode case, and the results of total three cases are reported in Table 1, the fiber parameters of which are the same to the ones introduced in Section 2.1. We use 1000 testing samples in every case to perform $M^2$ estimation and the reported PE and consuming time is the averaged value of these testing samples. In addition, the patterns are directly set as the corresponding resolution, so that the consuming time doesn't include the acquisition and transfer from the camera. It can be found that a higher resolution can achieve smaller prediction error but the processing time becomes relatively longer.

Table 1. Comparison of 3 different resolution cases investigated.

| Resolution | PE | Consuming time (GPU) |
|---|---|---|
| 96 × 96 | 1.79% | 4.86 ms |
| 128 × 128 | 1.64% | 5.14 ms |
| 160 × 160 | 1.53% | 5.87 ms |

Based on the results above, we can find that our scheme is accurate, robust and very fast for $M^2$ estimation. Compared with traditional methods [9–21], our approach only needs a single-shot near-field beam image collected by CCD to perform $M^2$ estimation, which is very economic and easy to implement under normal condition. Due to the computing efficiency of trained CNN, it only takes about 5ms to perform $M^2$ estimation for a single input pattern, which is a breakthrough in $M^2$ measuring efficiency. Moreover, different from traditional methods, our scheme has great anti-noise ability, which will be helpful for practical applications. We believe our approach is a novel and vital supplementary for $M^2$ determination methods.

### 3.2. Experimental demonstration

The experimental setup to verify the feasibility of our method is shown in Fig. 9. To give a reference standard of the $M^2$ parameter, the MD and VCM technique is adopted to measure the $M^2$ value [18], which has been proven to accurately agree with ISO-standard measurements but be capable to acquire substantial testing samples due to time efficiency. We adopt a pig-tailed single-frequency laser at 1064 nm as the light source. The delivery fiber is a single-mode fiber (SMF). The output end of the SMF is placed on a three-axis nano-position stage and the light from the SMF is coupled into the step-index few-mode fiber (FMF) with a core diameter of 25 μm and NA of 0.065. The V-value of this FMF is 4.80 at 1064 nm, thus supporting six eigenmodes. By adjusting the nano-stage, the coupling condition between the SMF and FMF can be changed so that near-field beam from the end facet of the FMF can be varied. The near-field pattern is imaged on Camera 1 (1920 × 1200 pixels of 5.86 μm pitch) through a 4-f imaging system consisting of a microscopic objective (MO) and an achromatic doublet lens (L1) whose focal lengths are 4 mm and 500 mm respectively. Accordingly, the magnification factor of the system is approximately 125. A polarization beam splitter (PBS) is located between L1 and Camera 1 to select only one polarization component of the beam and the beam splitter (BS) is to split beam for the measurement of $M^2$.

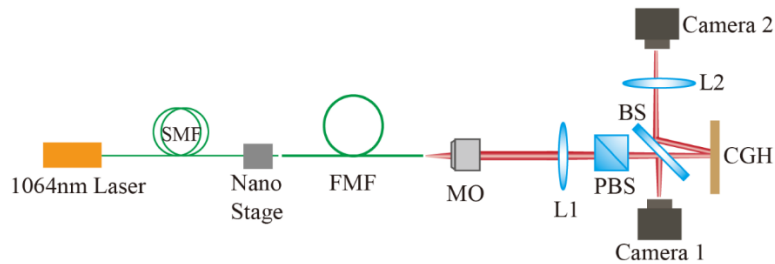

Fig. 9. Scheme of the experimental setup. SMF, single-mode fiber; FMF, few-mode fiber; L1, L2, lenses; MO, microscopic objective; BS, beam splitter; PBS, polarizing beam splitter; CGH, computer-generated hologram. The near-field beam pattern is recorded by Camera 1. The ground truth of $M^2$ is measured based on the VCM technique with the help of CGH.

To measure the reference value of $M^2$, a computer-generated hologram (CGH) is displayed on the spatial light modulator (SLM) with 1920 × 1080 pixels of 6.4 μm pitch. The light field is diffracted by the CGH and then Fourier transformed by a lens (L2) with focal length of 175 mm. Camera 2 is placed at the Fourier plane of the SLM to acquire the first rank diffraction pattern to acquire MD results [33]. Based on the MD results, the near-field is reconstructed and the intensities of the beam at different propagation length are simulated using the free-space transfer function. Then the $M^2$ value is determined based on the second-order moments of the beam at different length [18].

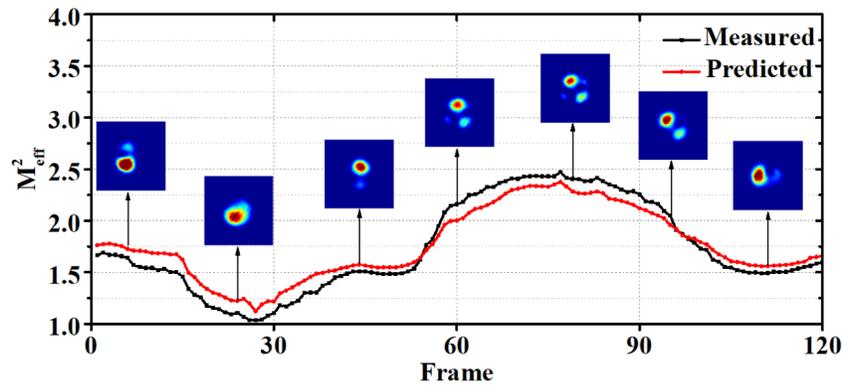

Fig. 10. The comparison of the measured $M^2$ and the predicted value by the pre-trained CNN. Some corresponding recorded near-field beam patterns are also shown in the insets.

By randomly adjusting the nano-stage, we acquire 120 varying near-field beam patterns recorded by Camera 1. Then these frames are cropped to make the patterns in the center of images and passed through the corresponding trained CNN for $M^2$ prediction. The predicted $M^2$ value and the corresponding measured one are compared in Fig. 10. It can be seen that the predicted value fluctuates in a small range around the measured one, which indicates the accuracy of our scheme. Some recorded patterns of certain frames are also displayed in the insets of Fig. 10 and the diversity of patterns shows the robustness of our approach. A video to show the measured near-field beam patterns as well as the measured and predicted $M^2$ is illustrated in Visualization 6. The calculated averaged PE for these 120 pattern samples is 6.32%, which is a bit larger than the simulated case, but is acceptable when there is not a very strict precision demand for $M^2$ estimation. As for the time cost, the total time to process all the pattern samples is 1.65 s, which means one pattern consumes about 14 ms to determine $M^2$. Noted the time is longer than the simulated case, the reason is that cropping the collected pattern to suitable size consumes relatively longer time. However, to the best of our knowledge, this is the fastest method to measure $M^2$. The extraordinary time efficiency indicates that our scheme can achieve well performance to determine $M^2$ in real time. Therefore, it can be utilized for online monitoring of $M^2$ in fiber laser manufacturing, fiber components evaluation, mode instabilities [34] research and other related cases.

## 4. Conclusion

In summary, an $M^2$ prediction scheme utilizing CNN has been proposed and its performance has been evaluated by simulated patterns for the first time as far as we know. Our scheme can achieve averaged prediction error smaller than 2% even when there are 10 modes involved in the fiber. In addition, this CNN can give $M^2$ prediction for one prepared beam pattern in only about 5 ms, indicating real-time ability with high performance. The patterns suffering strong noise can still be processed by our scheme, indicating great robustness. Besides, we give a brief discussion about the relation between the image resolution and the accuracy as well as time efficiency. The experiment results also prove the feasibility of our scheme. All in all, the superior performances of the novel $M^2$ evaluation method are endued by deep learning. We believe that this new scheme for estimating $M^2$ factor of fiber beams holds great promises for practical applications. Furthermore, the proposed method is not limited to the fiber beams as shown in this contribution, but can be applicable to other laser beams such as Laguerre-Gaussian beams and Hermite-Gaussian beams provided that adequate training samples are accessible.

## Funding

National Natural Science Foundation of China (NSFC) (61605246, 61805280); Research Grants from College of Advanced Interdisciplinary Studies, National University of Defense Technology (JC18-07).

## Acknowledgments

We thank Dr. Kun Xie for providing help in the experimental test of our method.

## References

1. A. E. Siegman, "New developments in laser resonators," Proc. SPIE **1224**, 2–14 (1990).
2. M. W. Sasnett and T. J. Johnston, "Beam characterization and measurement of propagation attributes," Proc. SPIE **1414**, 21–32 (1991).
3. H. Weber, "Some historical and technical aspects of beam quality," Opt. Quantum Electron. **24**(9), S861–S864 (1992).
4. A. E. Siegman, "How to (maybe) measure laser beam quality," in *Diode Pumped Solid State Lasers: Applications and Issues* (Optical Society of America, 1998), paper MQ1.
5. P. Zhou, Z. Liu, X. Xu, Z. Chen, and X. Wang, "Beam quality factor for coherently combined fiber laser beams," Opt. Laser Technol. **41**(3), 268–271 (2009).
6. Y. Ke, C. Zeng, P. Xie, Q. Jiang, K. Liang, Z. Yang, and M. Zhao, "Measurement system with high accuracy for laser beam quality," Appl. Opt. **54**(15), 4876–4880 (2015).
7. P. Yan, X. Wang, M. Gong, and Q. Xiao, "Evaluating the beam quality of double-cladding fiber lasers in applications," Appl. Opt. **55**(23), 6145–6150 (2016).
8. Y. Du, G. Feng, H. Li, Z. Cai, H. Zhao, and S. Zhou, "Real-time determination of beam propagation factor by Mach–Zehnder point diffraction interferometer," Opt. Commun. **287**, 1–5 (2013).
9. International Organization for Standardization, *ISO 11146 Test methods for laser beam widths, divergence angles and beam propagation ratios* (ISO,Geneva, 2005).
10. C. Schulze, D. Flamm, M. Duparré, and A. Forbes, "Beam-quality measurements using a spatial light modulator," Opt. Lett. **37**(22), 4687–4689 (2012).
11. R. D. Niederriter, J. T. Gopinath, and M. E. Siemens, "Measurement of the $M^2$ beam propagation factor using a focus-tunable liquid lens," Appl. Opt. **52**(8), 1591–1598 (2013).
12. R. Cortés, R. Villagómez, V. Coello, and R. López, "Laser beam quality factor ($M^2$) measured by distorted fresnel zone plates," Rev. Mex. Fis. **54**(4), 279–283 (2008).
13. M. Scaggs and G. Haas, "Real time laser beam analysis system for high power lasers," Proc. SPIE **7913**, 791306 (2011).
14. Y. Du, Y. Fu, and L. Zheng, "Complex amplitude reconstruction for dynamic beam quality $M^2$ factor measurement with self-referencing interferometer wavefront sensor," Appl. Opt. **55**(36), 10180–10186 (2016).
15. Z.-G. Han, L.-Q. Meng, Z.-Q. Huang, H. Shen, L. Chen, and R.-H. Zhu, "Determination of the laser beam quality factor ($M^2$) by stitching quadriwave lateral shearing interferograms with different exposures," Appl. Opt. **56**(27), 7596–7603 (2017).
16. S. Pan, J. Ma, R. Zhu, T. Ba, C. Zuo, F. Chen, J. Dou, C. Wei, and W. Zhou, "Real-time complex amplitude reconstruction method for beam quality $M^2$ factor measurement," Opt. Express **25**(17), 20142–20155 (2017).
17. O. A. Schmidt, C. Schulze, D. Flamm, R. Brüning, T. Kaiser, S. Schröter, and M. Duparré, "Real-time determination of laser beam quality by modal decomposition," Opt. Express **19**(7), 6741–6748 (2011).
18. D. Flamm, C. Schulze, R. Brüning, O. A. Schmidt, T. Kaiser, S. Schröter, and M. Duparré, "Fast $M^2$ measurement for fiber beams based on modal analysis," Appl. Opt. **51**(7), 987–993 (2012).
19. H. Yoda, P. Polynkin, and M. Mansuripur, "Beam quality factor of higher order modes in a step-index fiber," J. Lit. Technol. **24**(3), 1350–1355 (2006).
20. G. Bai, X. Chen, Y. Yang, Y. Zheng, X. Zhao, K. Liu, C. Zhao, Y. Qi, B. He, and J. Zhou, "Beam quality evaluation of 20/400 μm large-mode-area fiber based on mode decomposition and reconstruction," Laser Phys. **28**(2), 025101 (2018).
21. L. Huang, S. Guo, J. Leng, H. Lü, P. Zhou, and X. Cheng, "Real-time mode decomposition for few-mode fiber based on numerical method," Opt. Express **23**(4), 4620–4629 (2015).
22. K. Simonyan and A. Zisserman, "Very deep convolutional networks for large-scale image recognition," https://arxiv.org/abs/1409.1556.
23. K. He, X. Zhang, S. Ren, and J. Sun, "Deep residual learning for image recognition," in *Proceedings of the IEEE Conference on Computer Vision and Pattern Recognition* (IEEE, 2016), pp. 770–778.
24. J. Li, R. Klein, and A. Yao, "A two-streamed network for estimating fine-scaled depth maps from single rgb images," in *Proceedings of the 2017 IEEE International Conference on Computer Vision* (IEEE, 2017), pp. 3392–3400.
25. Y. LeCun, Y. Bengio, and G. Hinton, "Deep learning," Nature **521**(7553), 436–444 (2015).
26. D. Chen, S. Zhang, W. Ouyang, J. Yang, and Y. Tai, "Person Search via A Mask-Guided Two-Stream CNN Model," in *European Conference on Computer Vision* (Springer International Publishing, 2018), pp. 764–781.

27. S. W. Paine and J. R. Fienup, "Machine learning for improved image-based wavefront sensing," Opt. Lett. **43**(6), 1235–1238 (2018).
28. N. Borhani, E. Kakkava, C. Moser, and D. Psaltis, "Learning to see through multimode fibers," Optica **5**(8), 960–966 (2018).
29. A. Liu, T. Lin, H. Han, X. Zhang, Z. Chen, F. Gan, H. Lv, and X. Liu, "Analyzing modal power in multi-mode waveguide via machine learning," Opt. Express **26**(17), 22100–22109 (2018).
30. Y. An, L. Huang, J. Li, J. Leng, L. Yang, and P. Zhou, "Learning to decompose the modes in few-mode fibers with deep convolutional neural network," Opt. Express **27**(7), 10127–10137 (2019).
31. A. W. Snyder and J. Love, *Optical Waveguide Theory* (Springer Science & Business Media, 2012).
32. M. N. Zervas and C. A. Codemard, "High power fiber lasers: a review," IEEE J. Sel. Top. Quantum Electron. **20**(5), 219–241 (2014).
33. T. Kaiser, D. Flamm, S. Schröter, and M. Duparré, "Complete modal decomposition for optical fibers using CGH-based correlation filters," Opt. Express **17**(11), 9347–9356 (2009).
34. C. Jauregui, J. Limpert, and A. Tünnermann, "High-power fibre lasers," Nat. Photonics **7**(11), 861–867 (2013).